\documentclass[%
 aip,
 amsmath,amssymb,
 reprint,%
]{revtex4-1}

\usepackage{graphicx}
\usepackage{dcolumn}
\usepackage{bm}

\usepackage[utf8]{inputenc}
\usepackage[T1]{fontenc}
\usepackage{mathptmx}
\usepackage{braket}
\usepackage{amssymb}
\usepackage{siunitx}
\begin{document}

\preprint{AIP/123-QED}

\title[]{Quantum dots as potential sources of strongly entangled photons for quantum networks}

\author{Christian Schimpf}
\author{Marcus Reindl}
    \affiliation{Institute of Semiconductor and Solid State Physics, Johannes Kepler University, Linz 4040, Austria}
\author{Francesco Basso Basset}
    \affiliation{Department of Physics, Sapienza University, 00185 Rome, Italy}
\author{Klaus D. Jöns}
    \affiliation{Department of Physics, Paderborn Universiy, 33098 Paderborn, Germany}
\author{Rinaldo Trotta}
    \affiliation{Department of Physics, Sapienza University, 00185 Rome, Italy}
\author{Armando Rastelli}
    \affiliation{Institute of Semiconductor and Solid State Physics, Johannes Kepler University, Linz 4040, Austria}

\date{\today}

\begin{abstract}
The generation and long-haul transmission of highly entangled photon pairs is a cornerstone of emerging photonic quantum technologies, with key applications such as quantum key distribution and distributed quantum computing. However, a natural limit for the maximum transmission distance is inevitably set by attenuation in the medium. A network of quantum repeaters containing multiple sources of entangled photons would allow to overcome this limit. For this purpose, the requirements on the source's brightness and the photon pairs' degree of entanglement and indistinguishability are stringent. Despite the impressive progress made so far, a definitive scalable photon source fulfilling such requirements is still being sought for.
Semiconductor quantum dots excel in this context as sub-poissonian sources of polarization entangled photon pairs. In this work we present the state-of-the-art set by GaAs based quantum dots and use them as a benchmark to discuss the challenges to overcome towards the realization of practical quantum networks.
\end{abstract}

\maketitle


\section{Introduction}

After decades of fundamental research, quantum entanglement emerged as a pivotal concept in a variety of fields, such as quantum computation\cite{OBrien2009}, -communication\cite{Gisin2007,Kimble2008,Ekert1991} and -metrology \cite{Degen2017}. A manifold of quantum systems are beeing investigated and photons stand out in many areas due to their robustness against environmental decoherence and their compatibility with existing optical fiber\cite{Xiang2019} and free-space\cite{Yin2020} infrastructure. Non-local correlations were demonstrated in several photonic degrees of freedom, such as time-bin\cite{Simon2005, Jayakumar2014}, time-energy\cite{Tanzilli2005}, orbital angular momentum \cite{Mair2001}, polarization\cite{Aspect1981}, spin-polarization\cite{Bock2018,Vasconcelos2020} or in a combination of them ("hyper-entanglement"\cite{Kwiat1997,Prilmuller2018}). In quantum information processing the manipulation and measurement of entangled qubits plays a major role. Applications like quantum key distribution (QKD) with entangled qubits \cite{Ekert1991,Bennett1992,Acin2007} require high source brightness, high degree of entanglement, transmission through a low-noise quantum channel and finally a straightforward measurement at remote communication partners, all with minimal losses. These prerequisites could be met by polarization entangled photon pairs\cite{Wengerowsky2019}. Besides the most prominent sources based on spontaneous parametric down-conversion (SPDC)\cite{Kwiat1995,Giustina2015,Shalm2015}, also semiconductor quantum dots (QDs)\cite{Akopian2006,Gurioli2019,Huo2013,Huber2018,Reimer2012,Juska2013} are capable of generating polarization entangled photon pairs with a fidelity to a maximally entangled state above $0.97$\cite{Huber2018}. The probabilistic emission characteristics of SPDC sources prohibit a high brightness in combination with a high degree of entanglement\cite{Orieux2017}. This is not the case for QDs due to their sub-poissionian photon statistics \cite{Schweickert2018}. Furthermore, in a real-world context, applications like QKD require entanglement to be communicated over large distances\cite{Xiang2019,Yin2020} to be practically relevant. Most transmission channels, like optical fibers, underlie damping (about \SI{0.2}{dB/km} for common telecom-fibers), which severely limits the transmission range. This limitation can be alleviated by exploiting a concept of quantum communication\cite{Gisin2007}: The interconnection of multiple light sources in quantum networks \cite{Gisin2007,Kimble2008} via the realization of a cascaded quantum repeater scheme with entangled photons and quantum memories\cite{Briegel1998,Lloyd2001,Kimble2008}. In order to reach this goal, properties of the photon sources beyond the maximum entanglement fidelity become relevant, such as the photon indistinguishability \cite{Mandel1987}.
In this work, we examine the key figures of merit of entangled photon pairs with an emphasis on the distribution of entanglement in a quantum network. We will start from the state-of-the-art focusing on GaAs QDs. Although the emission wavelength of about \SI{785}{nm} is currently non-ideal for efficient fiber-based applications, GaAs QDs represent an excellent model system for the here discussed ideas due to their performance. Finally, we will outline recent concepts and approaches towards the realization of a viable quantum network.


\section{Polarization entangled photon pairs from quantum dots}

\begin{figure*}[ht]
\centering
\includegraphics[width=16.5cm]{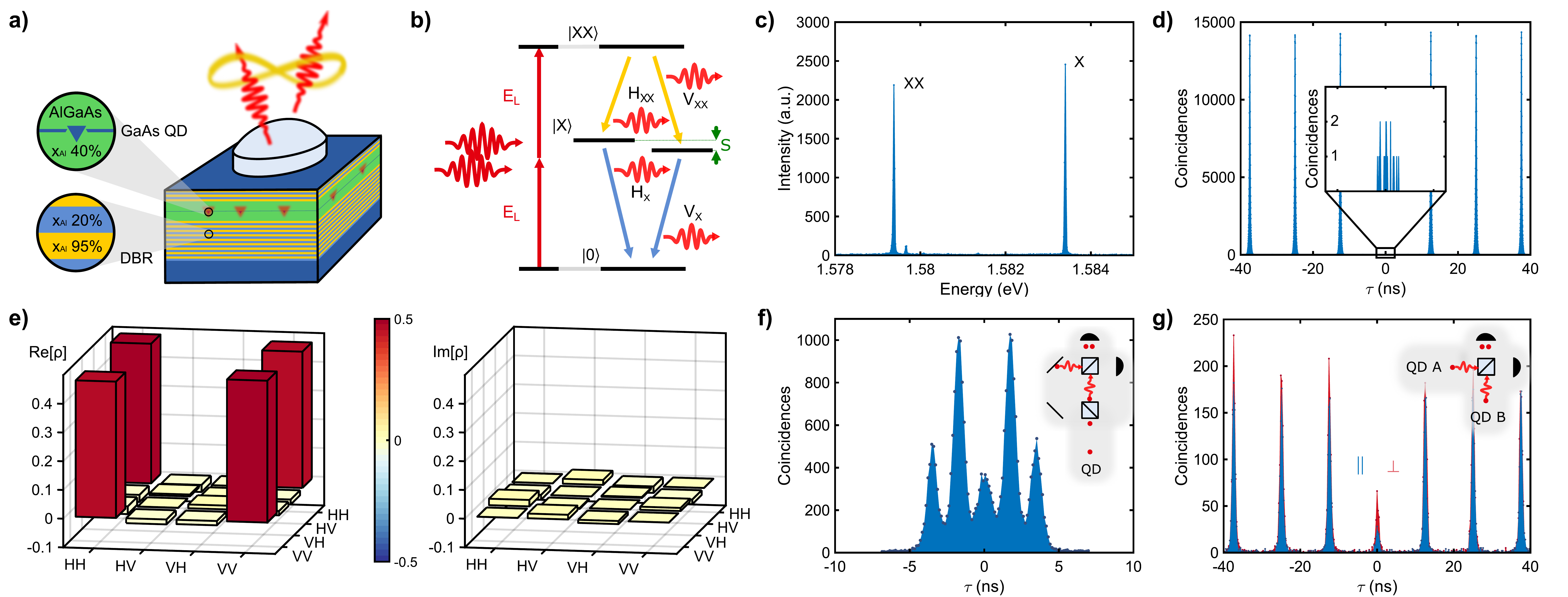}
\caption{\textbf{Compilation of measurements for GaAs QDs.} (a) Common sample structure with GaAs QDs in a lambda-cavity sandwiched between distributed Bragg reflectors (DBRs). In combination with a solid immersion lens (SIL) this yields an extraction efficiency of about 0.11. (b) Scheme of entangled photon pair generation using the resonant two-photon excitation (TPE) process. (c) Emission spectrum of under TPE. (d) Autocorrelation of the XX signal from a GaAs QD excited by TPE and measured by superconducting nanowire single photon detectors (SNSPDs), with a resulting $g^{(2)}(0)=\SI{7.5(16)e-5}{}$. Reproduced with permission from Ref. \citenum{Schweickert2018}. (e) Real- and imaginary part of the two-qubit density matrix of the X and XX in the horizontal (H) and vertical (V) polarization basis. The derived fidelity is $f_{\ket{\phi^+}}=0.97(1)$. (f) Two-photon interference visibility from one doubly excited QD with a time delay of \SI{2}{ns}. (g) Two-photon interference visibility for two remote QDs with a resulting interference visibility of $V=\SI{0.51(5)}{}$. Reproduced with permission from Ref.\citenum{Reindl2017}.}
\label{fig:GaAs_QD}
\end{figure*}

A common scheme to generate entangled photon pairs with semiconductor QDs embedded in photonic strucures (see Fig. \ref{fig:GaAs_QD}(a)) is by resonantly populating the biexciton (XX) state by a two-photon excitation (TPE) process \cite{Stufler2006}. The XX radiatively decays via the biexciton-exciton(X)-ground state cascade \cite{Benson2000}, as depicted in Fig. \ref{fig:GaAs_QD}(b). Ideally, the emitted photon pairs are in the maximum entangled Bell state $\ket{\phi^+}=1/\sqrt{2}\left( \ket{HH}+\ket{VV}\right)$, with $\ket{H}$ and $\ket{V}$ the horizontal and vertical polarization basis states, respectively. The fidelity $f_{\ket{\phi^+}}$ of the real photon pair's state to $\ket{\phi^+}$ is mostly determined by the fine structure splitting (FSS) $S$ and lifetime $T_\text{1,X}$ of the intermediate X state. In the absence of other dephasing mechanisms, the maximum achievable fidelity of an ensemble of photon pairs to a maximum entangled state is given by\cite{bassobasset2020arxiv}
\begin{equation}
    f^{\text{max}}_{\ket{\phi^+}}=\frac{1}{4}\left(2-g_0^{(2)}+\frac{2\left(1-g_0^{(2)}\right)}{\sqrt{1+\left( S\,T_\text{1,X} / \hbar \right)^2}} \right),
\label{eq:fmax}
\end{equation}
where $g_0^{(2)}$ is the multi-photon emission probability.
In the case of GaAs QDs obtained by the Al droplet etching technique \cite{Gurioli2019}, a high in-plane symmetry\cite{Huo2013} results in average FSS values below \SI{5}{\micro eV}, while the weak lateral carrier confinement \cite{Huber2019} causes radiative lifetimes of about $T_\text{1,XX}=\SI{120}{ps}$ and $T_\text{1,X}=\SI{270}{ps}$. The wavelength of the emitted light hereby lies around \SI{780}{nm} (see Fig. \ref{fig:GaAs_QD}(c)), with the XX and X photons separated by about \SI{2}{nm} (\SI{4}{meV}), allowing them to be split by color. In contrast to SPDC-based sources\cite{Jons2017} the pair-generation probability $\epsilon$ and the $g_0^{(2)}$ of QDs are decoupled due to the sub-poissonian emission characteristics\cite{Benson2000}. This led to demonstrated values of $g_0^{(2)}= \SI{8(2)e-5}{}$, with $\epsilon=0.9$ (measured via cross-correlation measurements between XX and X\cite{Wang2019}) under pulsed TPE\cite{Schweickert2018}, as illustrated by the corresponding auto-correlation measurement in Fig. \ref{fig:GaAs_QD}(d). Figure \ref{fig:GaAs_QD}(e) displays a resulting two-photon density matrix $\rho$ in the polarization space of the XX and X photons for an as-grown GaAs QD with $S\approx\SI{0.4}{\micro eV}$, acquired by full-state tomography\cite{James2001} under pulsed TPE at an excitation rate of $R=\SI{80}{MHz}$. The fidelity deduced from this matrix as $f_{\ket{\phi^+}}=\braket{\phi^+|\rho|\phi^+}$ is \SI{0.97(1)}. By utilizing a specifically designed piezo-electric actuator\cite{Trotta2015}, capable of restoring the in-plane symmetry and erasing the FSS of the QDs by strain, fidelity values up to \SI{0.978(5)}{} were demonstrated\cite{Huber2018}. These results suggest that a modest Purcell enhancement of a factor 3 could alleviate remaining dephasing effects and push the fidelity up to \SI{0.99}{}, which would match with the best results from SPDC sources\cite{Jons2017}.
The minimum time delay $1/R$ between the pulses depends on the lifetimes $T_\text{1,XX}$ and $T_\text{1,X}$, allowing for excitation rates up to $R\approx\SI{1}{GHz}$ (without Purcell enhancement). This makes GaAs QDs a viable source for applications like QKD with entangled photons \cite{Ekert1991,Bennett1992,Acin2007,Jennewein2000}, where high source brightness and near-unity entanglement fidelity are required in order to reach practical secure key rates in the high MHz or even GHz range.
In order to reach this goal, however, a high photon-pair extraction efficiency is required, which is naturally limited in semiconductor structures due to total internal reflection at the air/semiconductor interface. A simple approach to increase the pair extraction efficiency $\eta_E^{(2)}$ from less than $10^{-4}$ to about 0.01 is to embed the QDs in a lambda cavity defined between two distributed Bragg reflectors and adding a solid immersion lens on top\cite{Huber2016}, see Fig. \ref{fig:GaAs_QD}(a). Recently, circular Bragg resonators have demonstrated values of $\eta_E^{(2)}=\SI{0.65(4)}{}$\cite{Liu2019} and Purcell enhancement up to a factor $11.3$\cite{Wang2019}. Although a non-ideal entanglement fidelity due to the high FSS was reported in Ref. \citenum{Liu2019}, these structures are compatible with the aforementioned strain tuning techniques, which could cancel the FSS to create a bright source of highly entangled photon pairs, applicable for QKD with key rates potentially in the GHz range.

A widely discussed and researched topic is the distribution of entanglement over basically arbitrary distances, for which sources operating at high pair emission rates are especially relevant. One approach is free-space transmission via satellites, where recently a distance of \SI{1120}{km} was covered \cite{Yin2020}. From the practical point of view it would be desirable to exploit the existing and well-established telecom optical fiber networks\cite{Xiang2019}. The obvious effect of fibers on the transmitted light is a uniform damping, which is about $\SI{0.2}{dB/km}$ for typical fibers in the telecom C-Band (\SI{1550}{nm} wavelength). When transmitting polarization entangled photons through optical fibers, however, also polarization mode dispersion\cite{Antonelli2011} (PMD) has to be taken into account. PMD causes the principal states of polarization (PSPs) of the photons' wave packets to drift apart in time, leading to a degradation of the entanglement. The latter is proportional to the ratio of the total induced PMD $\tau$ between the two entangled photons and the length of the photon wave packets, given by $2\,T_1$. The maximum achievable fidelity to a perfectly entangled state, derived from the 2-qubit density matrix in polarization space\cite{Antonelli2011}, is then given by:
\begin{equation}
    f_{\text{PMD}}(\tau)= \frac{1}{2}+\left( \frac{1}{2} + \frac{\tau}{4\,T_{1}}\right)e^{\frac{-\tau}{2T_{1}}}.
\end{equation}
For simplicity, we assume that $T_1=\text{min}(T_{\text{1,X}},T_{\text{1,XX}})$, which represents a worst case scenario. If the two entangled photons experience exactly the opposite relative drift due to a maximum mismatch of the input modes with the PSPs, the PMD from the individual fibers add up to $\tau=\tau_1+\tau_2$. A typical value for the PMD of a single mode fiber is $D=\SI{0.1}{ps/\sqrt{km}}$ (e.g. the type "SMF-28e+" from "Corning"), thus with the given lifetimes and a length of $l=\SI{200}{km}$ for each the X and the XX photon's fibers, the maximum possible fidelity for $T_1=\SI{120}{ps}$ is still $f_{\text{PMD}}(2D\sqrt{l})>0.99$. For $T_1=\SI{10}{ps}$, the $f_{\text{PMD}}$ degrades to about $0.98$. For $T_1=\SI{1}{ps}$, which is approximately the case for sources based on the SPDC process, the maximum fidelity drops to about $0.79$, unless resorting to lossy measures like frequency filtering~\cite{Lim2016} or conversion to time-bin entanglement~\cite{Sanaka2002,Jayakumar2014}. If the input polarization modes are aligned with the PSPs, the total PMD reads $\tau=\tau_1-\tau_2$. Therefore, given that $\tau_1=\tau_2$, a configuration exists which exactly cancels out the effect of PMD and preserves the entanglement. This could be achieved by realigning the input modes to the PSPs during operation by polarization controllers\cite{Simon1990}.

An additional effect to consider in the context of a fiber-based network is chromatic dispersion, which leads to a temporal broadening of the wavefunctions\cite{Lenhard2016,Weber2019}. This lowers the success probability of two-photon interference, which we will discuss in the next section.


\section{Quantum dots in a quantum repeater based network}
Although transport of highly polarization entangled photons through fibers is possible\cite{Wengerowsky2020}, the exponential damping will inevitably lead to an insufficient qubit rate. A possible solution to this problem is the realization of a quantum repeater scheme. The common approach is based on the DLCZ protocol (and its variants), that relies on spin-photon entanglement~\cite{Duan2001}. However, the probabilistic nature of the entangling scheme limits the entanglement creation~\cite{Riebe2004}. Although an improved, deterministic version of the spin-photon based scheme was developed\cite{Humphreys2018}, the achieved rates are still modest. An alternative scheme relies on the use of entangled photon pair sources, like QDs, interfaced with quantum memories capable of receiving and storing entangled states to increase the qubit rate~\cite{Lloyd2001}. This scheme relies on a cascade of entanglement swapping processes \cite{Kirby2016,BassoBasset2019} among entangled photon pairs emitted by independent emitters. The teleportation of the entanglement is enabled by two-photon interference to perform a so-called Bell state measurement (BSM).
The success of a BSM strongly depends on the photon indistinguishability, which, in turn, depends on the photon sources and can be experimentally accessed by probing the interference visibility in a Hong-Ou-Mandel (HOM) experiment\cite{Lenhard2016}. For maximum visibility the spatio-temporal wave packets of the two photons involved in the BSM have to be identical and pure, i.e. no other physical system should contain information about the photon's origin. The latter point plays a crucial role in the case of QDs exploiting the decay cascade for entangled photon generation. The XX and X photons are correlated in their decay times\cite{Simon2005,Huber2013,schll2020crux}, which limits the maximum possible indistinguishability for both the XX and X photons to
\begin{equation}
    V_{\text{casc}}=\frac{1}{1+T_{\text{1,XX}}/T_{\text{1,X}}}
\label{eq:Vcasc}
\end{equation}
Figure \ref{fig:GaAs_QD}(f) shows the result of a HOM experiment with two XX photons generated by a GaAs QD, upon excitation with two pulses separated by a $\SI{2}{ns}$ delay. The resulting visibility of $0.69$ is typical for both the XX and X photons under TPE\cite{Huber2016} and is close to the maximum according to Eq. \eqref{eq:Vcasc}. As a comparison: For single X photons generated by resonant excitation a visibility of over $0.9$ is achieved for the same QDs\cite{Scholl2019,Reindl2019}.
When interfacing two dissimilar QDs, the inherently stochastic nature of the epitaxial growth has to be considered, which primarily leads to varying emission energies. Further, imperfections in the solid-state environment of the QDs lead to inhomogeneous broadening due to charge noise\cite{Kuhlmann2013}. This results in a jitter of the central emission energy by  $\delta E$ (full width at half maximum) around a mean value at time-scales typically in the microsecond to millisecond timescale \cite{Schimpf2019}. The jitter leads to a degradation of the indistinguishability described by\cite{Kambs2018}:
\begin{equation}
    V_{\delta E}=\frac{\hbar\,\text{Re}[\text{F}(z)]}
    {\sqrt{8\pi}\,\sigma\,T_1},
\label{eq:VE}
\end{equation}
with $\sigma=\delta E / 2\sqrt{2\text{ln}2}$ the standard deviation of the Gaussian distribution of the energy jitter and $\text{Re}[\text{F}(z)]$ the real part of the Faddeeva function of $z=i\,\hbar/(2\,\pi\,\sqrt{2}\,\sigma\,T_1)$.
Figure \ref{fig:GaAs_QD}(g) shows the result of a HOM experiment with two X photons from GaAs QDs with $\delta E\approx \SI{4}{\micro eV}$ and a resulting visibility of $V={0.51(5)}$ \cite{Reindl2017}. The center wavelengths were previously matched by tuning the energy of one QD via a monolithic piezo-electric actuator. 

\begin{figure*}[ht]
\centering
\includegraphics[width=16.5cm]{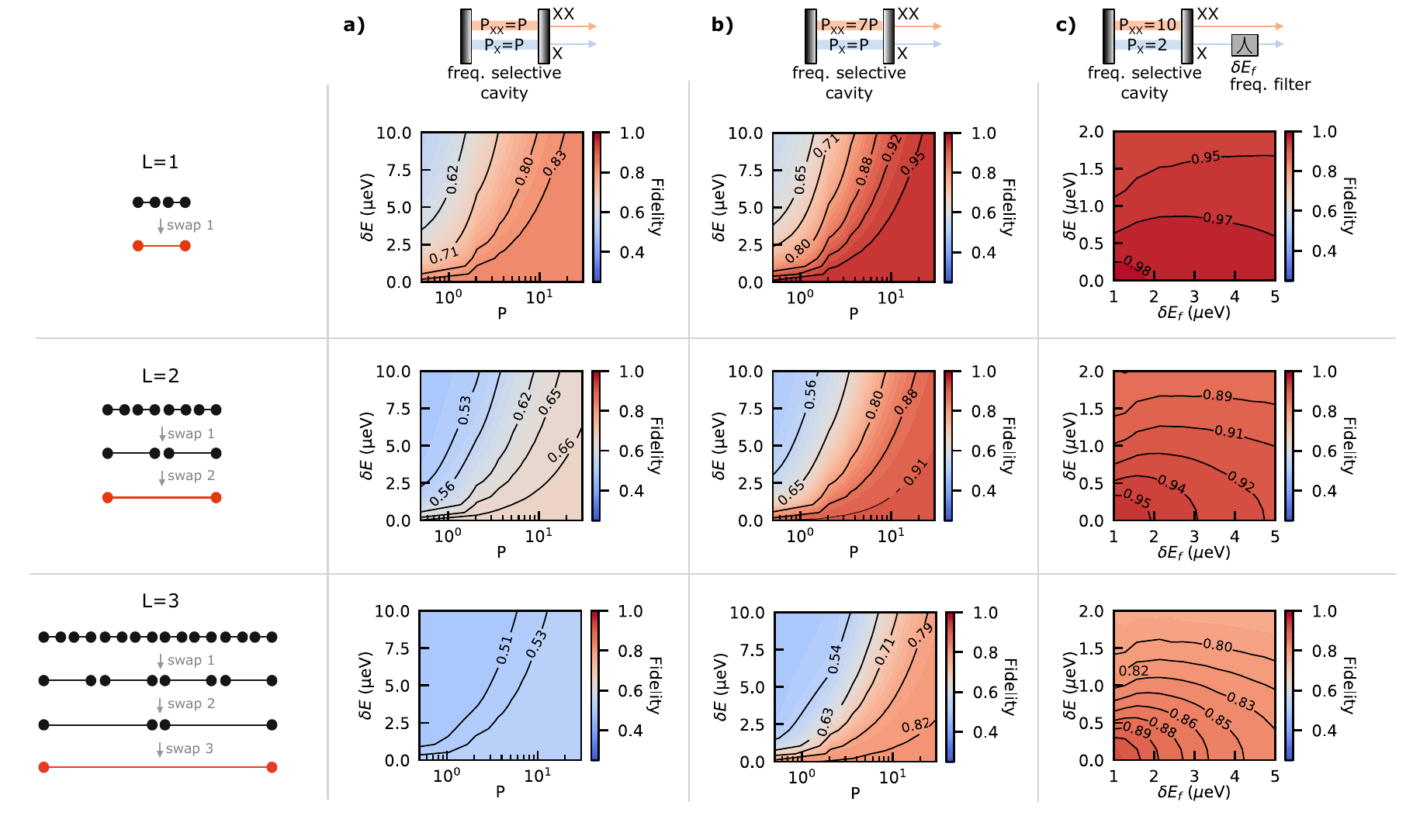}
\caption{\textbf{Entanglement fidelity in a chain of quantum relays.} Simulated entanglement fidelity of the final entangled photon pair in a chain of quantum relays performing entanglement swapping operations among pairs of polarization entangled photons emitted by QDs under TPE. The chain depths are $L=\{1,2,3\}$. All QDs are assumed to have an FSS of $\SI{0.05}{\micro eV}$. (a) Fidelity as a function of Purcell factor $P$ and Gaussian energy jitter $\delta E$. (b) Fidelity with a frequency selective cavity, so that $P_{\text{X}}=P$ and $P_{\text{XX}}=7P$ . (c) Fidelity as a function of the Lorentzian width of a frequency filter $\delta E_f$ and the Gaussian energy jitter $\delta E$ (full width at half maximum), for fixed Purcell factors of $P_{X}=2$ and $P_{XX}=10$.}
\label{fig:Fidelity}
\end{figure*}

We will now discuss possible solutions to overcome the two major indistinguishability degrading mechanisms in QDs discussed so far: The partial temporal entanglement in the XX-X decay cascade (Eq.\eqref{eq:Vcasc}) and frequency jitter (Eq.\eqref{eq:VE}). Both effects are influenced by the radiative lifetimes $T_{\text{1,XX}}$ and $T_{\text{1,X}}$, which can be modified by exploiting the Purcell effect in a cavity\cite{Liu2019,Wang2019}. Figure \ref{fig:Fidelity} illustrates concatenated entanglement swapping processes with a depth of $L=\{1,2,3\}$, i.e. a chain of quantum relays forming the backbone of quantum repeaters. The number of QDs required is $2^L$, while the range covered scales with $2^L\,l_0$, with $l_0$ the total length of both fibers departing from one QD. This example serves as a demonstration how the entanglement fidelity evolves over multiple layers of swapping operations with photons generated by QDs.
Fig. \ref{fig:Fidelity}(a) depicts the final entanglement fidelity as a function of the Purcell factor $P$ and the energy jitter $\delta E$. In all simulations the natural linewidth of the exciton is set to \SI{2.4}{\micro eV} for $P=1$, corresponding to $T_{\text{1,X}}=\SI{270}{ps}$. Values of $P>15$ are unpractical, as the total relaxation time of the QD then approaches the typical excitation pulse width of about $\SI{10}{ps}$. This primarily leads to an increasing re-excitation probability\cite{Gustin2018,Hanschke2018}, which is detrimental to the indistinguishability and the entanglement. In addition, PMD effects in optical fibers start to become relevant for such short wave packets. For the calculation of the fidelity we utilize the density matrix formalism for describing one entanglement swapping process with QDs\cite{BassoBasset2019} with a type of BSM which can detect two Bell states\cite{Mattle1996,bassobasset2020arxiv} ($\psi^+$ and $\psi^-$). In order to model a chain of entanglement swapping processes the formalism is applied recursively, assuming uncorrelated BSM success probabilities in successive steps. We simultaneously account for varying lifetimes caused by $P$ and a decreased BSM success rate due to $\delta E$ (see supplementary for details). From the simulations we observe that already for two swapping processes the homogeneous Purcell enhancement alone cannot recover the entanglement fidelity sufficiently, as it merely alleviates the impact from inhomogeneous broadening on the indistinguishability, but the visibility degrading effect from the XX-X cascade is still at full force. Figure \ref{fig:Fidelity}(b) depicts the case for an energy selective cavity\cite{Huber2013}, which enhances the XX decay rate by a factor of 7 compared to the X, so that $P_{X}=P$ and $P_{XX}=7P$. This approach could strongly increase the BSM success rate and therefore the resulting entanglement fidelity. However, the finite temporal width of the excitation pulse, whose minimum value is set by the  the limited spectral separation between X and XX and the necessity of suppressing laser stray light, sets a lower limit to the lifetimes - and therefore an upper limit to the Purcell enhancement - in order to limit re-excitation \cite{Gustin2018}. A compromise could be achieved by mild frequency filtering of the X photon, as illustrated in Fig. \ref{fig:Fidelity}(c). Filtering partially erases the temporal information held by the X photon, leading to the same outcome as prolonging the X lifetime and hence decreasing the XX/X lifetime ratio as with the selective Purcell enhancement. In the simulations we assume a filter with Lorentzian transmission characteristics and a FWHM of $\delta E_f$ and an energy jitter with a FWHM of $\delta E$ for both the X and the XX (see supplementary for details). We assume a frequency selective cavity with fixed Purcell enhancement of $P_\text{X}=2$ and $P_\text{XX}=10$. As a result of the filtering, the effective lifetime of the X signal increases while simultaneously reducing the impact of the energy jitter. Note that for the here investigated values of $\delta E$ the interference visibility again drops for $\delta E_f$ values below the inhomogeneous broadening $\delta E$. In addition, in the presence of a finite FSS, the BSM success rate drops when the filtered linewidth is in the order of the FSS or below\cite{BassoBasset2019}. From the simulations we can observe that with a low inhomogeneous broadening ($<\SI{0.2}{\micro eV}$) and a moderate frequency filtering of about $\SI{4}{\micro eV}$ one could achieve an entanglement fidelity of approximately 0.93 at $L=2$ and 0.85 at $L=3$.

A complete repeater scheme requires also quantum memories~\cite{Lvovsky2009} which can store and retrieve a photonic qubit with high fidelity. To address the noise and bandwidth limitation of quantum memories two groups invented a cascaded absorption memory scheme, which is intrinsically noise-free.~\cite{Kaczmarek2018,Finkelstein2018} Furthermore the possibility to use an off-resonant Raman transition in this cascaded scheme allows for large storage bandwidth, limited mainly by the available control laser power. Currently the main drawback of these schemes is the limited storage time, which is determined by the radiative lifetime of the upper state of the cascade (below \SI{100}{ns}). Another promising approach is to use rare-earth doped crystals as quantum memories~\cite{Afzelius2009}, featuring performances that equalize, if not outperform, those of cold atomic ensembles~\cite{Li2016,Cao2020} or trapped emitters in terms of efficiency~\cite{Hedges2010} and coherence times~\cite{Zhong2015}. These memories have shown a full quantum storage protocol with telecom-heralded quantum states of light~\cite{Seri2017}, and the first photonic quantum state transfer between nodes of different nature~\cite{Maring2017}. Furthermore, atomic frequency comb quantum memories were the first to be successfully interfaced with single photons emitted from a quantum dot~\cite{Tang2015}.


\section{Future outlook}

We conclude that bright and nearly on-demand sources of highly entangled photon pairs are on the verge of becoming reality. The groundwork has been laid through the development of semiconductor quantum dots (QDs) emitting highly entangled photons\cite{Huber2018}, of advanced optical cavity structures\cite{Liu2019,Wang2019} and technology capable of manipulating the symmetry and emission energy of QDs\cite{Trotta2012,Trotta2015}. On-chip integration of QDs\cite{Dietrich2016} and the implementation of electric excitation schemes\cite{Salter2010} can further increase the practicability in emerging quantum technology.

The optimal wavelength (about \SI{1550}{nm}) for transporting entangled photons through fibers is currently determined by the established telecom fiber infrastructure. Material systems to obtain QDs emitting at this wavelength are under development\cite{Olbrich2017,Xiang2019,zeuner2019arxiv} and existing sources with emission at shorter wavelengths could be adapted by frequency down-conversion\cite{Weber2019}. Recently, a basic GHz-clocked quantum relay with QDs emitting directly in the Telecom-C band was demonstrated\cite{Anderson2020}.
One of the greatest, yet rewarding challenges is the interfacing of dissimilar sources of entangled photons for multi-photon applications\cite{Vural2020} and long-haul entanglement distribution\cite{Xiang2019,Yin2020}. The physical limits to the indistinguishability\cite{Simon2005,schll2020crux} set by the currently employed cascade for entangled photon pair generation\cite{Benson2000,Huber2018} and fluctuations stemming from the solid state environment of QDs\cite{Huo2013,Kuhlmann2013} pose intricate challenges for the years to come. As demonstrated in this work, the application of selective Purcell enhancement together with mild frequency filtering could alleviate the limit of indistinguishability of the entangled photon pairs.
The different emission energies and radiative lifetimes of the biexciton (XX) or exciton (X) in QDs could be matched by utilizing strain-\cite{Trotta2015} and electric\cite{Trotta2013} degrees of freedom independently. Considering the quantum relay chains depicted in Fig. \ref{fig:Fidelity}, three strain degrees of freedom can cancel the fine structure splitting and adapt the central energy of the XX or X to the next neighbor's. The electric degree of freedom can simultaneously be used to fine-tune the respective radiative lifetime and therefore the shape of the photonic wave-packet. By repeating this strategy through the whole relay chain for each QD, one could optimize the resulting entanglement fidelity of the final photon pair.

With these tools at hand, the next leap towards the demonstration of a functional quantum network will be interconnection of two dissimilar quantum dots via entanglement swapping \cite{Kirby2016,BassoBasset2019}. The following steps could be to interface the photons performing the Bell state measurement with quantum memories\cite{Lvovsky2009,Kaczmarek2018,Afzelius2009,Hedges2010,Zhong2015,Seri2017,Maring2017,Tang2015} and use the resulting entangled photon pairs for quantum key distribution\cite{Ekert1991,Bennett1992,Acin2007} with efficiencies beating the direct transmission through fibers. From there on, the goal is to expand the system to a chain of multiple QDs and implement a quantum repeater scheme  \cite{Briegel1998,Lloyd2001,Kimble2008} in order to enhance the resulting entanglement fidelity and efficiency compared to a repeater-less distribution scheme.

\section*{Acknowledgement}
Christian Schimpf is a recipient of a DOC Fellowship of the Austrian Academy of Sciences at the Institute of Semiconductor Physics at Johannes Kepler University, Linz, Austria.
This project has received funding from the Austrian Science Fund (FWF): FG 5, P 29603, P 30459, I 4380, I 4320, I 3762, the Linz Institute of Technology (LIT) and the LIT Secure and Correct Systems Lab funded by the state of Upper Austria and the European Union's Horizon 2020 research and innovation program under grant agreement No. 820423 (S2QUIP), No. 899814 (Qurope) and No. 679183 (SQPRel).

\section*{Data Availability}
Data sharing is not applicable to this article as no new data were created or analyzed in this study.

\section*{References}
\bibliography{main}

\end{document}